\documentclass{article}    

\usepackage[pdftex]{graphicx} %for arXiv
\usepackage{amsfonts}
\usepackage{amssymb}
\usepackage{amsmath}
\usepackage{mathrsfs}
\usepackage[subrefformat=parens]{subcaption}

\newcommand{\bm}[1]{\mbox{\boldmath{$#1$}}} 
\newcommand{\be}{\begin{eqnarray}}
\newcommand{\en}{\end{eqnarray}}
\newcommand{\ben}{\begin{eqnarray*}}
\newcommand{\enn}{\end{eqnarray*}}

\newcommand{\lb}{\left(\begin{array}{@{\,}c@{\,}}}
\newcommand{\lbc}{\left(\begin{array}{@{\,}c@{\,}}}
\newcommand{\rb}{\end{array}\right)}
\newcommand{\lbcc}{\left(\begin{array}{@{\,}cc@{\,}}}
\newcommand{\lbccc}{\left(\begin{array}{@{\,}ccc@{\,}}}
\newcommand{\lbcccc}{\left(\begin{array}{@{\,}cccc@{\,}}}
\newcommand{\lbccccc}{\left(\begin{array}{@{\,}ccccc@{\,}}}
\newcommand{\lbcccccc}{\left(\begin{array}{@{\,}cccccc@{\,}}}
\newcommand{\lbccccccc}{\left(\begin{array}{@{\,}ccccccc@{\,}}}
\newcommand{\lbcccccccc}{\left(\begin{array}{@{\,}cccccccc@{\,}}}
\newcommand{\case}{\left\{\begin{array}{ll}}
\newcommand{\encase}{\end{array} \right.}

\title{Heterogeneous source model for magnetoenecephalography
}

\author{Takaaki Nara$^1$, Ten-yu Yang$^1$, 
and Kenta Kabashima$^1$\\
$^1$ The University of Tokyo
}

\begin{document}

\maketitle

\section*{Abstract.} 
In this paper, we propose a novel source model for a magnetoencephalography (MEG) inverse problem 
that combines a conventional extended parametric approach and an imaging approach. 
Our aim is to separately identify a focal current source and background activities spread over the brain. 
The new source model consists of two terms to represent different spatial characteristics: 
one is a localized patch source represented with a few parameters based on a mapping from a sphere to the cortex surface, 
and the other is a distributed source expressed using elemental dipoles on grid points on the cortical surface.  
We call it a heterogeneous source model, because these two models have not been used simultaneously. 
Effectiveness of the proposed method is shown via numerical simulations. 

\section{Introduction}

Magnetoencephalography (MEG) is a noninvasive monitoring tool for brain activity
that is widely used for analysis of brain functions and medical diagnosis. 
In particular, localization of an epileptic focus for ablative surgery is one of crucial applications of MEG, 
since 
synchronized, strong currents flowing in the focus can be inversely estimated from the magnetic field measured outside a patient's head. 
Among the noninvasive modalities for brain activities, 
MEG has an advantage in that it has high temporal resolution 
because the magnetic field generated by the neural currents can be regarded as quasi-static. 
However, reconstruction of the currents with high spatial resolution is a challenge 
due to the ill-posed nature of an inverse source problem using MEG. 
For its realization, use of a source model that constrains a solution 
 based on  {\it a priori} physiological knowledge is essential.  

Source models in the MEG inverse problem 
are categorized into two groups \cite{Baillet01, Grech08, Becker15}: 
an equivalent current dipole (ECD) model used in parametric approaches 
and a source model using elemental dipoles distributed over a cortical surface in imaging approaches. 
The first category is used when the neural activity to be estimated is focal 
and hence it is assumed that it is represented by a single ECD. 
When multiple focal activities are expected to exist, 
a model assuming a few ECDs is also used. 
Under these models, 
the positions and moments of the ECDs are obtained 
by nonlinear optimization \cite{Huang98, Uutela98}, 
scanning methods such as multiple signal classification (MUSIC) \cite{Mosher92, Mosher99_RAPMUSIC}, 
adaptive spatial filtering \cite{vanVeen97, Vrba01, Sekihara05}, 
or an algebraic method \cite{Nara07}. 
By contrast, the second category is used when neural activities are spread over the brain. 
In this model, the current distribution is represented by elemental dipoles fixed on grids 
on the cortical surface, the moments of which are obtained 
by solving an under-determined inverse problem with some regularization. 
For details, see \cite{Grech08} and references therein.  
Both methods have pros and cons: 
the former is suitable for focal current sources and 
can accurately identify the centers of the localized activities, 
but it cannot identify the spatial extent of the sources. 
In addition, the estimated ECD positions are affected by background activities 
spread over the cortical surface which are not well modeled by ECDs. 
By contrast, the latter can represent the distribution of current sources, 
but the obtained result for a localized source tends to be blurred when using L2-norm regularization 
or scattered around the true source 
even when using sparse regularization such as L1-norm and total variation regularization. 
Also, all these imaging approaches strongly depend on choice of regularization parameters. 
Moreover, it is often difficult to separate a focal source of main interest 
from background activities distributed over the cortical surface. 

To identify a spatial extent of a focal source, 
several methods in the middle of the two approaches
have been proposed. 
They are further categorized into two groups. 
The first approach represents a focal source 
parametrically up to its spatial extent, 
which we call in this paper an extended parametric approach.
Lutkenhoner {\it et al.} proposed expressing the extended source 
by a patch source, which is a uniformly activated 
cortical area giving rise to distributed currents which flow perpendicular to the cortical surface \cite{Lutkenhoner95}. 
Kincses {\it et al.} \cite{Kincses99}
proposed a method that begins by adjusting several dipoles 
on the cortical surface and then expands them by adding neighboring dipoles 
to minimize the residual of data. 
They also 
proposed a method to represent a patch approximately 
by a circular patch on the cortical surface parametrically 
in terms of its center, radius, and current density \cite{Kincses03}. 
Then, the parameters for $N$ patches are obtained 
by maximizing the likelihood using an algorithm 
in which a random walk for the seed locations with a random radius  is performed. 
David {\it et al.} \cite{David02} proposed a time-coherent expansion method 
to estimate the spatial extent of cortical areas of time-coherent activity. 
Yetik {\it et al.} \cite{Yetik06} expressed a patch in terms of a parametric surface, 
including a part of a spherical surface as a special case, 
and then maximum and minimum values of its parameters were obtained 
to determine the patch. 
An interesting method to represent extended sources parametrically 
was proposed by Im {\it et al.} \cite{Im05} in which a one-to-one correspondence between 
the cortical surface and a sphere is used. 
Expressing the patch sources in terms of the parameters of bell-shaped functions 
defined on the sphere, they are searched by the gradient-based method. 
Haufe {\it et al.} \cite{Haufe11} modeled the current density 
as a linear combination of Gaussian bases, 
and estimated their center positions, the variances, and the amplitudes. 

The second category to identify spatial extent of a focal source is based on extension of the scanning methods for the ECD source model. 
Limpiti {\it et al.} \cite{Limpiti06} expanded the current density in a patch 
using a set of local basis functions  
and treating its expansion coefficients as unknown parameters. 
Hiilbrand {\it et al.} \cite{Hillebrand11} modified a nonlinear minimum variance beamformer (SAM \cite{Vrba01}) 
for a circular patch source. 
Birot {\it et al.} \cite{Birot11} proposed to estimate spatially-extended sources 
by union of circular shaped sources on a cortical surface, 
where the circular domains were selected
based on the MUSIC-like metric computed by using the $2q$-th order $(q\ge 1)$ statistical matrix of the data. 
They also proposed the source model composed of epileptic activities and background activities. 
Based on the assumption that the processes of the epileptic activities are not Gaussian 
whereas those of the background activity are Gaussian, 
they are separated using $2q$-th order cumulants. 
Becker {\it et al.} \cite{Becker14b, Becker15} proposed a method to identify extended sources 
by disk selection based on a different metric 
after the background activities are separated by a tensor-based preprocessing technique for a time-series data.

In this paper, we consider a problem 
to separately identify a single focal source, such as an epileptic focus, 
which is the main interest of identification,
and other background activities 
using single time shot data. 
In this case, the position and shape of the focal domain should be estimated 
as accurately as possible in the presence of background activities spread over the cortical surface. 
For that purpose, the conventional methods described above have significant problems.  
%that it is difficult to separate the focal source of interest  
%from the background activities, using a single time shot data. 
Using either of the ECD model, the distributed dipole model, or the extended parametric method,
as long as the current sources are represented by a single type of  source model, 
it is difficult to decompose the estimated sources into a focal and background activities. 
For example, when using an extended parametric method in \cite{Im05} based on 
the mapping between the cortical surface and a sphere, 
both of a focal and background activities are expressed 
by several domains mapped from the bell-shaped functions on a sphere, 
so that they are hardly distinguished. 
Although the methods in \cite{Limpiti06, Birot11, Becker14b} assume both the focal and background activities, 
they are separated by time-series data which degrades temporal resolution. 

In this study, we develop a method to combine an extended parametric approach with a patch source model 
and an imaging approach with a source model using elemental dipoles 
to separately obtain both the focal source and the background activities. 
We call our model a heterogeneous source model for a single time shot data, 
because two different kinds of models are included simultaneously. 
As for the extended parametric approach in the heterogeneous model, 
we use a mapping from the sphere to the cortical surface 
as in Im's method \cite{Im05}. 
Here, we express a patch as an image of a circular domain 
on the sphere so that the patch is represented 
by three parameters, which are the coordinates of the center position 
and the radius of the circular domain on the sphere. 
To separate the focal source from the background activities, 
we set a cost function composed of a misfit term 
between the measured magnetic field data at a single time shot 
and a regularization term of an L2-norm of the background activity, 
which is shown to be optimized in terms of the focal source parameters. 

The rest of this paper is organized as follows. 
Section \ref{sec:Theory} describes our heterogeneous source model with which solution for an inverse problem to separately identify a focal source and background activities is derived. 
In section \ref{sec:Simulations}, the proposed method is verified via numerical simulations. 
%We compare an imaging approach with L1 and total variation (TV) regularization, an extended parametric approach, 
%and the proposed method with the heterogeneous source model. 
Conclusions are given in section \ref{sec:conclusion}.

\section{Theory}
\label{sec:Theory}

\subsection{Heterogeneous source model}

Assume that a mesh with $M$ nodes is set on a 
cortical surface $\Sigma$. 
Let $\bm{J} \in \mathbb{R}^M$ be a current source on $\Sigma$ 
whose $i$th component represents a current dipole moment at the $i$th node. 
Our heterogeneous source model is expressed as 
\be
\bm{J} = \bm{J}_p(\bm{\theta}) + \bm{J}_b ,
\label{eq:hetero}
\en
where $\bm{J}_p(\bm{\theta}) \in \mathbb{R}^M$ is a patch source 
whose position and shape are designated by the parameters $\bm{\theta}$ 
and $\bm{J}_b$ is a distributed elemental dipole source. 
To represent both the focal and background activities, 
the conventional parametric and extended parametric approaches assume $\bm{J}_p$ only, 
whereas the conventional imaging approach assumes $\bm{J}_b$ only. 
In contrast, our model assumes both of them and determines them separately.

\begin{figure}[htbp] 
\begin{center}
\includegraphics[width=10cm]{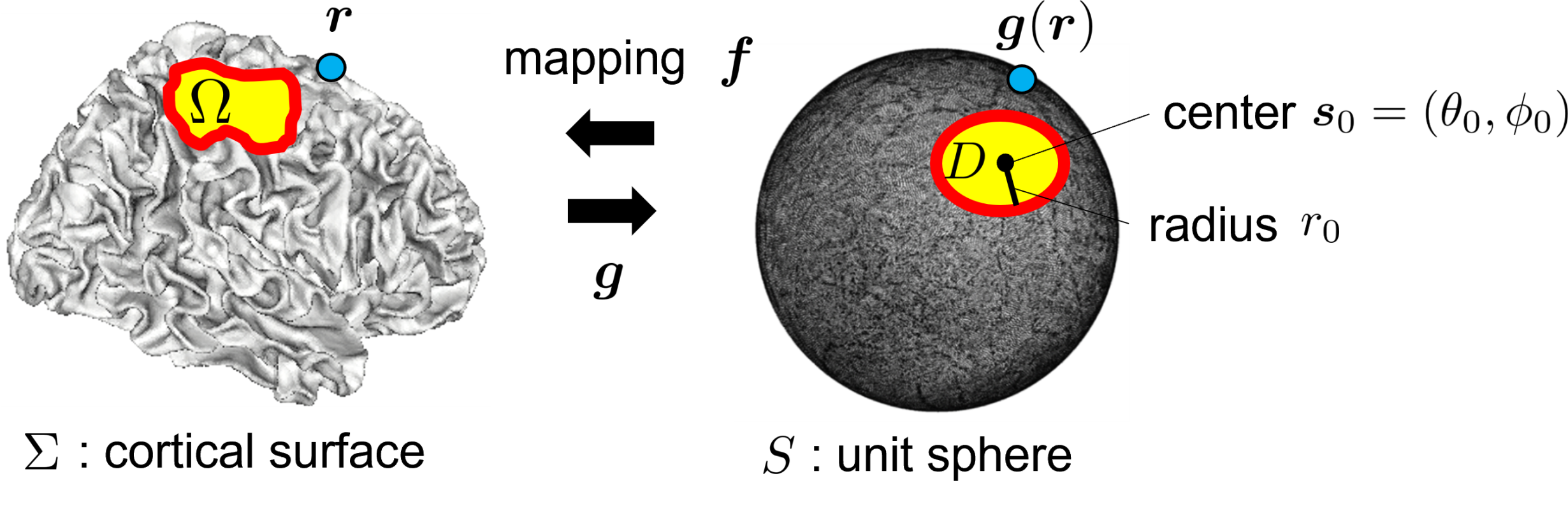}
\end{center}
\caption{A mapping between a cortical surface and a sphere}
\label{fig:mapping}
\end{figure}

To express the position and shape of a patch source on the cortical surface with a few parameters, 
following Im \cite{Im05}, we use a mapping from a sphere to the cortical surface. 
Let $\bm{g}$ be a map from the cortical surface $\Sigma$ to the unit sphere denoted by $S$, 
and let $\bm{f}$ be its inverse, 
as shown in Fig.~\ref{fig:mapping}. 
Then, a focal domain $\Omega \subset \Sigma$ 
can be regarded to be an image of a simply connected domain $D \subset S$ 
via $\bm{f}$ : $\Omega = \bm{f}(D)$. 
Numerical construction of the mapping $\bm{g}$ and $\bm{f}$ based on a subject's MR image 
has been proposed so far; see, for example, 
\cite{Zeng13} 
and references therein. It is also implemented in Freesurfer. 
In this paper, we assume that $D$ is a circular domain on $S$. 
Although $\Omega$ expressed under this assumption is limited, 
the position and the shape of a focal domain on $\Sigma$
can be represented in terms of only three parameters: 
$\bm{\theta} = ( \theta_0, \phi_0, r_0)$, where 
$(\theta_0, \phi_0)$ are the spherical coordinates of
the center position of $D$ 
and $r_0$ is the radius of $D$. 
Also, following the extended parametric model in \cite{Kincses03}, 
we assume that the current density $j_0$ in $\Omega$ is unknown but constant. 
The method can be easily generalized without difficulty 
to the case where it has a distribution 
expressed as a basis expansion \cite{Limpiti06} 
or another function shape, such as a bell-shaped function \cite{Im05} 
or Gaussian.

Under these assumptions, 
we express a focal source whose current density is $j_p$ homogeneously in $\Omega$ and zero in $\Sigma / \Omega$. 
First, it is expressed using a continuous function on $\Sigma$ by 
\be
J_p(\bm{r}) 
= j_p H( r_0 - d(\bm{g}(\bm{r}), \bm{s}_0)) ,
\quad \bm{r} \in \Sigma ,
\en
where $\bm{g}(\bm{r}) \in S$ is the point mapped from $\bm{r} \in \Sigma$, 
$\bm{s}_0 \equiv (\sin\theta_0 \cos\phi_0, \sin\theta_0 \sin\phi_0, \cos\theta_0)$ 
is the center position of $D$ expressed with Cartesian coordinates, 
$d(\bm{g}(\bm{r}), \bm{s}_0) \equiv \cos^{-1} (\bm{g}(\bm{r}) \cdot \bm{s}_0$) 
is the distance along $S$ 
between $\bm{g}(\bm{r})$ and $\bm{s}_0$, 
and $H$ is the Heaviside function (which is 1 when its argument is positive and 0 otherwise). 
When $\bm{r}$ exists in $\Omega$, 
the corresponding point $\bm{g}(\bm{r})$ is included in $D$, 
and hence the Heaviside function becomes 1 so that $J_p(\bm{r}) = j_p$, 
whereas when $\bm{r}$ is outside $\Omega$, $J_p(\bm{r}) = 0$. 
Now, for $M$ nodes at $\bm{r}_1, ..., \bm{r}_M$ on $\Sigma$, 
let 
\be
\bm{H}(\bm{s}_0, r_0)
\equiv
(H( r_0 - d(\bm{g}(\bm{r}_1), \bm{s}_0)), ..., 
H( r_0 - d(\bm{g}(\bm{r}_M), \bm{s}_0)) )^T
\in \mathbb{R}^M ,
\en
whose component is one or zero depending on 
whether the node exists in $\Omega$. 
Using this vector, we express a focal source on the cortical surface in a discretized form by 
\be
\bm{J}_p = j_0 \bm{H}(\bm{s}_0, r_0) \in \mathbb{R}^M, 
\label{eq:patch}
\en
where $j_0$ is a current dipole moment 
which is constant at the nodes inside $\Omega$.  
This is our parametric patch source model.

In addition to the focal source, 
we represent background activities spread over the whole cortical surface by $\bm{J}_b \in \mathbb{R}^M$ 
as in a usual imaging approach. 
Although the probability distribution that each component of $\bm{J}_b$ obeys can be arbitrarily chosen, 
in this paper, we assume that a distributed source 
is as a realization of the $M$th degree normal distribution: 
\be
\bm{J}_b \in \mathbb{R}^M
\sim
N(\bm{0}, \Sigma_b) ,
\label{eq:background}
\en
where $\Sigma_b = \sigma_b^2 I$. 

Eq.~(\ref{eq:hetero}) with Eqs.~(\ref{eq:patch}) and (\ref{eq:background}) constitute our heterogeneous source model. 
If $\bm{J}_p$ and $\bm{J}_b$ are obtained 
from a single time-shot magnetic field data, 
the focal patch source $\bm{J}_p$ and the background activities $\bm{J}_b$ spread over the cortical surface 
are separately identified. 

\noindent {\bf Remark.} The focal source in Eq.~(\ref{eq:patch}) includes not only $\bm{\theta} = (\bm{s}_0, r_0)$ but also $j_0$ as unknown parameters. However, as shown in section \ref{sec:Jp}, $j_0$ can be represented in terms of $\bm{\theta}$ as a solution to a linear least-squares problem. Hence, 
$\bm{\theta}$ is the substantial parameters in $\bm{J}_p$. 

\subsection{Solving an inverse problem with the heterogeneous source model}

\subsubsection{Solution for $\bm{J}_p$}
\label{sec:Jp}

We assume that there exist $N$ sensors, 
which are the magnetoemters, gradiometers, 
or their combination. 
Let $\bm{d} = (d_1, ..., d_N)^T$ be data 
at a single time shot 
and  $L \in \mathbb{R}^{N \times M}$ be a leadfield matrix.  
Then we have
\be
\bm{d}
=
L (\bm{J}_p + \bm{J}_b) + \bm{n} ,
\label{eq:observation}
\en
where $\bm{n}$ represents measurement noise 
and is assumed that $\bm{n} \sim N(\bm{0}, \Sigma_n)$ 
where $\Sigma_n = \sigma_n^2 I$. 
In Eq.~(\ref{eq:observation}),
because both $\bm{J}_b$ and $\bm{n}$ are assumed to obey 
a normal distribution,
$\bm{d} - L \bm{J}_p
=
L \bm{J}_b + \bm{n} 
$
also follows a normal distribution
$N(\bm{0}, \sigma_b^2 L L^T + \sigma_n^2 I)$. 
Thus, the likelihood function is given by
\be
p(\bm{d} | \bm{s}_0, r_0, j_0)
\propto
\exp( - ( \bm{d} - L \bm{H}(\bm{s}_0, r_0) j_0 )^T (\sigma_b^2 L L^T + \sigma_n^2 I)^{-1} ( \bm{d} - L \bm{H}(\bm{s}_0, r_0)  j_0 )) .
\label{eq:likelihood}
\en
Let us consider the maximization of Eq.~(\ref{eq:likelihood}).
First, 
assuming that $\sigma_b$ and $\sigma_n$ are given, 
for fixed $\bm{s}_0$ and $r_0$, 
the optimum $j_0$ is obtained by a linear least-squares method. 
Denoting it by $\hat{j}_0(\bm{s}_0, r_0)$, 
$\bm{J}_p$ can be regarded as the function of 
$\bm{s}_0$ and $r_0$ only as
$\bm{J}_p(\bm{s}_0, r_0) \equiv \hat{j}_0(\bm{s}_0, r_0) 
\bm{H}(\bm{s}_0, r_0)$. 
Using this, we set a cost function as
\be
\Phi(\bm{s}_0, r_0)
=
|| \bm{d} - L \bm{J}_p(\bm{s}_0, r_0) ||_{\Sigma^{-1}}^2 ,
\label{eq:cost}
\en
where $|| \bm{x} ||_A^2 = \bm{x}^T A \bm{x}$ and 
\be
\Sigma \equiv \sigma_b^2 L L^T + \sigma_n^2 I.
\label{eq:Sigma}
\en 
Here, the ranges of the unknown parameters are limited to
\be
\theta_0 \in [0, \pi],
\quad
\phi_0 \in [0, 2\pi],
\quad 
r_0 \in [0, r_{\max}] .
\label{eq:cube}
\en
For minimization of $\Phi(\bm{s}_0, r_0)$,  
where the unknown parameters are in the cube given by Eq.~(\ref{eq:cube}), 
an adaptive diagonal curve (ADC) method \cite{Sergeyev06} can be used, 
which is guaranteed to reach the global minimum of the cost function 
if it is Lipschitz continuous. 
To let the cost function in Eq.~(\ref{eq:cost}) be Lipschitz continuous, 
for numerical computation, 
we use a smeared-out Heaviside function,  
\be
\tilde{H}(\psi)
=
\case
0 , & \psi < - \epsilon \\
\frac{1}{2} + \frac{\psi}{2\epsilon} + \frac{1}{2\pi} \sin \frac{\pi \psi}{\epsilon} ,
& |\psi| \le \epsilon \\
1 , & \psi > \epsilon \\
\encase
\en
instead of the Heaviside function. 
$\epsilon$ is a fixed constant with an order of the side length of 
a mesh element. 
It is notable that, when the patches extended to opposite walls of sulci and gyri, 
 substantial cancellation of the generated magnetic field occurred \cite{Ahlfors10}. 
Hence, it is appropriate to set $r_{\max}$ such that a corresponding patch on the cortical surface 
does not spread over opposite walls of sulci and gyri. 
We also remark that, under the assumption that 
$\Sigma_b = \sigma_b^2 I$ and $\Sigma_n = \sigma_n^2 I$, Eq.~(\ref{eq:cost}) is rewritten as
\be
\Phi(\bm{s}_0, r_0)
=
\sigma_b^{-2}
( \bm{d} - L \bm{J}_p(\bm{s}_0, r_0) )^T
 (L L^T + (\frac{\sigma_n}{\sigma_b})^2 I)^{-1}
( \bm{d} - L \bm{J}_p(\bm{s}_0, r_0) ) .
\en
Hence, for minimization of $\Phi(\bm{s}_0, r_0)$, 
not each value of $\sigma_n$ and $\sigma_b$ 
but its ratio $\sigma_n/\sigma_b$ only is necessary. 

\subsubsection{Solution for $\bm{J}_b$}
\label{sec:Jb}

Once the minimizer $\hat{\bm{J}}_p$ of $\Phi$ in Eq.~(\ref{eq:cost}) is obtained, 
we consider the minimization problem
\be
\Psi(\hat{\bm{J}}_p, \bm{J}_b)
\equiv
|| \bm{d} - L (\hat{\bm{J}}_p + \bm{J}_b) ||_{\Sigma_n^{-1}}^2 
+
|| \bm{J}_b ||_{\Sigma_b^{-1}}^2
\rightarrow 
\min
\en
Because this is a simple linear inversion with L2-norm regularization, 
a unique solution is given by 
\be
\hat{\bm{J}}_b
=
(L^T \Sigma_n^{-1} L + \Sigma_b^{-1})^{-1} L^T \Sigma_n^{-1} ( \bm{d} - L \hat{\bm{J}}_p) .
\label{eq:Jbhat}
\en
This gives us an estimate of the background activities. 
When $\Sigma_b = \sigma_b^2 I$ and $\Sigma_n = \sigma_n^2 I$, Eq.~(\ref{eq:Jbhat}) is written as
\be
\hat{\bm{J}}_b
=
(L^T L + (\frac{\sigma_n}{\sigma_b})^2 I )^{-1} L^T ( \bm{d} - L \hat{\bm{J}}_p) ,
\label{eq:Jbhat2}
\en
which requires the ratio $\sigma_n/\sigma_b$ only again.

\subsubsection{Optimality of $\hat{\bm{J}}_p$ and $\hat{\bm{J}}_b$}

To examine the optimality of the solution obtained in sections \ref{sec:Jp} and \ref{sec:Jb}, 
we next consider the minimization problem
\be
\Psi(\bm{J}_p, \bm{J}_b)
\equiv
|| \bm{d} - L (\bm{J}_p(\bm{s}_0, r_0) + \bm{J}_b) ||_{\Sigma_n^{-1}}^2 
+
|| \bm{J}_b ||_{\Sigma_b^{-1}}^2
\rightarrow 
\min
\label{eq:Psi}
\en
This is a general form of a cost function 
composed of the noise-covariance-weighted squared error under the heterogeneous source model 
with a weighted L2-norm regularization term for the background activities. 
Here arises a question: 
does the two-step procedure in sections \ref{sec:Jp} and \ref{sec:Jb}, in which the squared error term is minimized first for $\bm{J}_p$ 
and then the total $\Psi$ is minimized for $\bm{J}_b$, 
give an optimum solution? 
In other words, is there a better combination of $\bm{J}_p$ and $\bm{J}_b$ that minimizes $\Psi$? 
To answer this question, let us note that, 
for an arbitrary fixed $\bm{J}_p$, 
the minimizer of $\bm{J}_b$ for $\Psi(\bm{J}_p, \bm{J}_b)$ is given by 
\be
\hat{\bm{J}}_b(\bm{J}_p)
=
(L^T \Sigma_n^{-1} L + \Sigma_b^{-1})^{-1} L^T \Sigma_n^{-1} ( \bm{d} - L \bm{J}_p) .
\label{eq:linearsol}
\en
Hence, minimization of $\Psi(\bm{J}_p, \bm{J}_b)$ for $\bm{J}_p$ and $\bm{J}_b$ 
is equivalent to minimization of $\Psi(\bm{J}_p, \hat{\bm{J}}_b(\bm{J}_p))$ for $\bm{J}_p$ with Eq.~(\ref{eq:linearsol}). 
However, we can prove that
\be
\Psi(\bm{J}_p, \hat{\bm{J}}_b(\bm{J}_p))
=
\Phi(\bm{J}_p) .
\en
%See the Appendix for the proof. 
Hence, minimization of  $\Psi(\bm{J}_p, \bm{J}_b)$ for $\bm{J}_p$ and $\bm{J}_b$ 
is equivalent to minimization of $\Phi(\bm{J}_p)$ for $\bm{J}_p$. 
Therefore, if we solve the minimization of $\Phi$ in Eq.~(\ref{eq:cost}) first for $\bm{J}_p$ 
and then substitute the obtained solution $\hat{\bm{J}}_p$ 
into Eq.~(\ref{eq:linearsol}), 
we obtain an optimum solution  
in the sense that they minimize $\Psi(\bm{J}_p, \bm{J}_b)$ in Eq.~(\ref{eq:Psi}).

\section{Numerical example}
\label{sec:Simulations}

In this section, a numerical example is shown 
that illustrates effectiveness of the proposed method. 
MRI data for an averaged cortical surface in \cite{Holmes09}
(Colin27) were used. 
The number of mesh elements of the left or right hemisphere was 331,025. 
A mapping between the cortical surface and a sphere was generated 
using FreeSurfer. Also, segmentation was conducted for the MRI data using FieldTrip 
to obtain a tissue-wise electrical conductivity map. 
In accordance with \cite{Haueisen95}, 
the values in Table \ref{table:sigma} are assigned for the tissues. 
With these conductivities, the boundary integral equation was solved 
using the linear quick Galerkin method \cite{Stenroos12} 
to obtain a lead field matrix.
As sensors, 204-channel gradiometers (Electa, Neuromag) were assumed. 

\begin{table}[htb]
\centering
  \caption{Electrical conductivity of tissues}
  \begin{tabular}{|l||c|}  \hline
    Tissue & Conductivity (S/m)  \\ \hline \hline
    Scalp & 0.4348 \\ \hline
    Skull & 0.00625 \\ \hline
    CSF & 1.5385 \\ \hline
    Grey matter & 0.3333 \\ \hline
    White matter & 0.1429 \\ \hline
  \end{tabular}
\label{table:sigma}
\end{table}

\begin{figure}[htbp]
  \begin{minipage}[b]{0.4\linewidth}
    \centering
    \includegraphics[width=4.5cm]{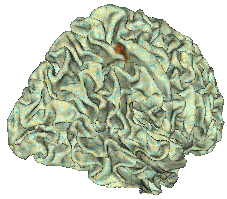}
    \subcaption{true}
    \label{true_fig1}
  \end{minipage}
  \begin{minipage}[b]{0.4\linewidth}
    \centering
    \includegraphics[width=4.5cm]{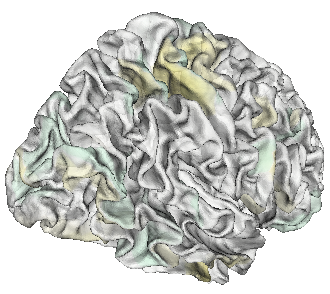}
    \subcaption{result of method (i)}
  \end{minipage}
  \begin{minipage}[b]{0.15\linewidth}
    \centering
    \includegraphics[width=1.5cm]{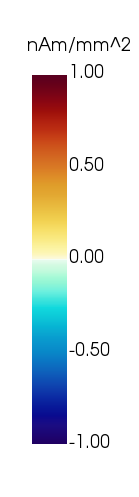}
  \end{minipage}\\
  \begin{minipage}[b]{0.4\linewidth}
    \centering
    \includegraphics[width=4.5cm]{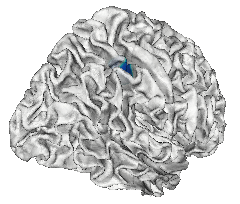}
    \subcaption{result of method (ii)}
    \label{conv_fig1}
  \end{minipage}
  \begin{minipage}[b]{0.4\linewidth}
    \centering
    \includegraphics[width=4.5cm]{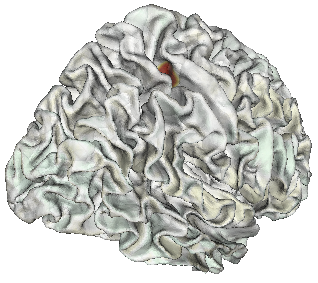}
    \subcaption{result of method (iii)}
  \end{minipage}
\caption{Reconstruction results}
\label{fig:example}
\end{figure}

A true patch source was generated by mapping a circular domain on $S$ with $(\theta_0, \phi_0, r_0) = (0.4 \, \hbox{rad}, -0.58 \, \hbox{rad}, 0.1 \, \hbox{rad})$. 
According to Murakami and Okada \cite{Murakami15}, the current dipole moment density in human neocortex was 
in the range of 0.16 to 0.77 nAm/mm$^2$. 
Following this, 
the current moment density $j_0$ was assumed to be 0.6~nAm/mm$^2$. 
The standard deviation $\sigma_b$ of the background activity was set to $\sigma_b / j_0 = 0.17$. 
The standard deviation of the measurement noise, $\sigma_n$, was given such that $\sigma_n / || L (\bm{J}_p + \bm{J}_b) || = 0.1$. 
The total source, consisting of the patch source and the background activities, is shown in Fig.~\ref{fig:example} (a). 
%In this subsection, $\sigma_b$ and $\sigma_n$ are assumed to be known.  
%Cases when they are unknown so that the true and the assumed values differ are examined in the next subsection. 
In this paper, we assume that 
$\sigma_b$ and $\sigma_n$ are known. 

We compared three methods: (i) imaging approach with L1 and TV regularization, 
(ii) extended parametric approach 
assuming only the patch source model given by Eq.~(\ref{eq:patch}), 
and (iii) proposed method 
assuming the heterogeneous source model 
given by Eq.~(\ref{eq:hetero}) with 
Eqs.~(\ref{eq:patch}) and (\ref{eq:background}). 
In method (i), following \cite{Becker14b}, we obtained the current distribution $\bm{J}$ by minimizing 
\be
\frac{1}{2}||L \bm{J} - \bm{d}||^2 + \lambda ( ||V \bm{J}||_1 + \alpha ||\bm{J}||_1)
\en
where $V$ represents a matrix for computing the total variation on the discretized cortical surface. 
The regularization parameter $\lambda$ was determined 
based on generalized cross validation 
where the ratio of the parameter for the TV term to that for the L1-norm term was fixed at $\alpha = 0.67$ 
according to the range $0.01 \le \alpha \le 1$ suggested in \cite{Becker14a}. 

Fig.~\ref{fig:example} (b), (c), and (d) shows the results for methods (i), (ii), and (iii), respectively. 
The source obtained by imaging approach (i) is not focal but instead scattered,  
making it difficult to clearly separate a focal source from the background activities. 
Although the extended parametric approach (ii) identifies a localized domain, the position of the patch deviates from that of the true one due to the effect of the background activities. 
This is because the model in method (ii) assumes a single patch only, and hence the obtained domain is a patch that equivalently represents the true patch source 
plus the background activities. 
In contrast to these two results, 
the proposed method (iii) can separately identify 
the patch source %(Fig.~\ref{fig:separate} (a)) 
and the background activities %(Fig.~\ref{fig:separate} (b)) 
%in the total source (Fig.~\ref{fig:example} (d)), 
where the estimated patch closely coincides with the true patch. 

%\begin{figure}[htbp]
%  \begin{minipage}[b]{0.35\linewidth}
%    \centering
%    \includegraphics[width=4.5cm]{fig/prop.png}
%    \subcaption{patch}
%  \end{minipage}
%  \begin{minipage}[b]{0.1\linewidth}
%    \centering
%    \includegraphics[width=1.5cm]{fig/bar.png}
%  \end{minipage}
%  \begin{minipage}[b]{0.35\linewidth}
%    \centering
%    \includegraphics[width=4.5cm]{fig/prop.png}
%    \subcaption{background activities}
%  \end{minipage}
%  \begin{minipage}[b]{0.1\linewidth}
%    \centering
%    \includegraphics[width=1.5cm]{fig/bar.png}
%  \end{minipage}
%\caption{Patch and background activities separately obtained by method (iii). 
%Note that the range of the color bar for (b) is different from that for (a) 
%to clearly see the weaker background activities.}
%\label{fig:separate}
%\end{figure}

\section{Conclusion}
\label{sec:conclusion}

In this paper, we proposed a heterogeneous source model for an MEG inverse problem 
by combining a conventional extended parametric approach and an imaging approach 
to separate a focal source from background neural activities. 
To represent a focal source, a patch model on the cortical surface is employed,  
which is expressed with three parameters based on a mapping from a sphere to the cortical surface. 
To express the distributed background activities, elemental dipoles on a grid are used. 
With this model, we proposed a two-step algorithm: first the parameters of the patch source are obtained 
using an optimization algorithm, the ADC method, which is guaranteed to converge to a global optimum. 
Second, the background activities are obtained by solving a linear inverse problem 
with L2-norm regularization for the background activities. 
It was shown that this algorithm gives an optimal solution 
that minimizes a cost function 
consisting of the squared error between the data and the magnetic field 
generated by the patch source and the background activities with an L2-norm regularization term 
for the background activities. 
A numerical example illustrated that the proposed method identified a focal patch more accurately 
in the presence of background activities than a conventional extended parametric approach or an imaging approach with L1 and TV regularization. 

\section*{Acknowledgement}
This paper was supported by JST PRESTO JPMJPR15E9 and JSPS KAKENHI 19H04438.

\bibliographystyle{IEEEtran}
\bibliography{MEG_patch.bib}

\end{document}